\documentclass[10pt,conference]{IEEEtran}

\IEEEoverridecommandlockouts
\usepackage{cite}
\usepackage{amsmath,amssymb,amsfonts}
\usepackage{algorithmic}
\usepackage{graphicx}
\usepackage{textcomp}
\usepackage{xcolor}
\usepackage{hyperref}
\usepackage{booktabs}
\usepackage{enumitem}
\usepackage{graphicx,booktabs}
\definecolor{baseline}{RGB}{0,100,0}
\definecolor{nobaseline}{RGB}{180,0,0}
\usepackage{xcolor}

\usepackage{setspace}
\usepackage{tabularx}  
\usepackage{booktabs}
\begin{document}

\title{Explaining Code Risk in OSS: Towards LLM-Generated Fault Prediction Interpretations}

\author{\IEEEauthorblockN{ Elijah Kayode Adejumo}
\IEEEauthorblockA{\textit{Computer Science} \\
\textit{George Mason University}\\
Fairfax, USA \\
eadejumo@gmu.edu}
\and
\IEEEauthorblockN{Brittany Johnson}
\IEEEauthorblockA{\textit{Computer Science} \\
\textit{George Mason University}\\
Fairfax, USA \\
johnsonb@gmu.edu}
}
\setstretch{0.93}

\maketitle

\begin{abstract}
Open Source Software (OSS) has become a very important and crucial infrastructure worldwide because of the value it provides. OSS typically depends on contributions from developers across diverse backgrounds and levels of experience.
Making safe changes, such as fixing a bug or implementing a new feature, can be challenging, especially in object-oriented systems where components are interdependent. Static analysis and defect-prediction tools produce metrics (e.g., complexity, coupling) that flag  potentially fault-prone components, but these signals are often hard for contributors new or unfamiliar with the codebase to interpret. Large Language Models (LLMs) have shown strong performance on software engineering tasks such as code summarization and documentation generation. Building on this progress, we investigate whether LLMs can translate fault-prediction metrics into clear, human-readable risk explanations and actionable guidance to help OSS contributors plan and review code modifications. We outline explanation types that an LLM-generated assistant
could provide (descriptive, contextual, and actionable explanations). 
We also outline our next steps to assess usefulness through a task-based study with OSS contributors, comparing metric-only baselines to LLM-generated explanations on decision quality, time-to-completion, and error rates.
\end{abstract}


\section{Introduction \& Background}
Contributions to open source software are crucial to its sustainability. 
Active contributor engagement and the on-boarding of newcomers are key components of an open source project. 
While open source projects often attract global contributions, diverse challenges arise as these audiences bring different levels of technical experience, language, and educational backgrounds \cite{steinmacher2015systematic,steinmacher2015social}. 
This can be especially problematic given the need to be able to understand project documentation, issues, and code to make meaningful contributions.



Open source projects face a unique challenge in that they rely heavily on newcomers who often lack deep codebase knowledge needed to make quality contributions. 
Fault proneness prediction tools~\cite{chidamber1994metrics, denaro2002empirical} can address this by providing metrics and binary classifications that reveal potentially faulty areas, enabling contributors to quickly understand code quality patterns without extensive project history. 
Numerous studies have demonstrated that structural and process metrics can effectively identify fault-prone code components \cite{subramanyam2003empirical,balasubramaniam2022software,goyal2022effective}. 
These predictive approaches offer significant practical benefits, including reduced testing effort, improved resource allocation, and informed architectural refactoring decisions. 

While such information could be effective in revealing and prioritizing areas for careful consideration and monitoring, a critical gap persists between tool availability and use in practice.
A recent study revealed that 68\% of fault prediction models provide no explanation of their output,leaving developers to independently interpret metric values and assess their relevance to specific project contexts~\cite{grattan2024need}. 
This ``explanation gap'' significantly hampers the adoption and effectiveness of defect prediction tools, as practitioners struggle to translate raw metric values (e.g.,"CBO=15") into actionable development decisions. 
Furthermore, lack of explanation behind tool notifications has been known to increase cognitive load, hinder trust, and reduce the likelihood of adoption and use~\cite{johnson2013don,dam2018explainable}.




Recent advances in Large Language Models (LLMs) have demonstrated remarkable capabilities in software engineering tasks, including code summarization \cite{ahmad2021transformer}, documentation generation and simplification~\cite{adejumo2024towards,adejumo2025bridging}, as well as bug report analysis \cite{tufano2024unveiling}.
Recent studies indicate that LLM-generated explanations and fault location suggestions can improve developer understanding while enabling faster task completion~\cite{kang2024quantitative,nam2024using}. 
Therefore, building on these findings, we hypothesize that LLM-generated explanations for fault prediction metrics could similarly improve trust and adoption of these tools. Our proposed approach combines established software risk metrics with the narrative capabilities of LLMs, addressing a crucial usability gap in automated static code analysis. 
We hypothesize that such LLM-powered explanations can improve contributors' ability to understand potential faultiness in open source software codebases, plan and review code changes, make more informed testing decisions, and ultimately lead to safer modifications and evolution of OSS projects. 





This paper makes the following contributions to address the explainability gap in OSS fault prediction tools:
\begin{enumerate}[label=\roman*.]
    \item \textbf{Explanation Taxonomy:}We propose a taxonomy 
    (descriptive, contextual,and actionable) specifically designed for 
    fault prediction metric explanations, grounded in developer information 
    needs from prior empirical studies.
    \item \textbf{Project-Aware Prompting Methodology:} We develop a 
    structured prompt composition approach that incorporates 
    project-specific statistical baselines (mean, standard deviation) 
    to ground LLM explanations in local context, addressing the 
    limitation of generic metric interpretations.

    \item \textbf{Feasibility Demonstration:} Through illustrative 
    scenarios on real OSS projects (Apache Ant 1.7 and Camel 1.6), 
    we demonstrate that LLMs can generate comprehensive explanations 
    that address all three taxonomy categories, converting raw CK 
    metrics into developer-friendly guidance.

\end{enumerate}

In the following sections, we present our proposed explanation taxonomy, describe how an LLM-based system could implement it, and outline plans to evaluate its effectiveness.
Our early illustrations shows that LLMs can bridge the gap between technical metrics and developer understanding, with the potential to transform fault proneness scores into accessible explanations and practical next steps for contributors across experience levels.




\section{Generating Explanations}
In this paper, we explore the ability for LLMs to generate comprehensive, human-readable explanations that can make fault prediction metrics more accessible to OSS contributors, particularly newcomers with diverse backgrounds and experience levels. 
We hypothesize that LLMs-generated \textit{descriptive}, \textit{contextual}, and \textit{actionable} explanations can improve developer understanding and task performance compared to traditional metric-only approaches.
Our efforts aim to answer the overarching question \textit{\textbf{To what extent can LLMs generate explanations for software fault prediction metrics that are descriptive, contextual, and actionable?}}
All artifacts, prompts, and outputs are publicly available for replication and reuse~\cite{ArtifactsCodeRisk}.

\subsection{Dataset Selection and Metric Extraction}
We conducted our preliminary exploration using Chidamber-Kemerer (CK) metrics from a publicly available software fault proneness prediction dataset~\cite{aggarwal2021software}. 
The Chidamber-Kemerer (CK) metrics suite represents one of the most influential contributions to this field, introducing object-oriented metrics such as Coupling Between Objects (CBO), Lack of Cohesion of Methods (LCOM), and Response for Class (RFC)~\cite{chidamber1994metrics}. Extensive empirical studies have validated the predictive power of these metrics across diverse software systems, establishing them as reliable indicators of fault-proneness in object-oriented codebases~\cite{basili2002validation}.
We sampled two projects: \textbf{\texttt{Apache Ant 1.7}} (745 classes; 166/22.3\% documented bugs),  and \textbf{\texttt{Apache Camel 1.6}} (965; 188/19.5\%), spanning build tooling, and enterprise integration.
For each project, we extracted the some CK metrics suite including Coupling Between Objects (CBO), Response for Class (RFC), Lack of Cohesion of Methods (LCOM), and Weighted Methods per Class (WMC), along with fault labels indicating historical bug occurrence. We computed project-specific baseline statistics to establish contextual thresholds shown in Table \ref{tab:project-metric-stats}.

\begin{table}[t]
\caption{Project-Specific Metric Statistics}
\label{tab:project-metric-stats}
\centering
\resizebox{\columnwidth}{!}{%
\begin{tabular}{lcccc}
\toprule
\textbf{Project} & \textbf{CBO ($\mu$, $\sigma$)} & \textbf{RFC ($\mu$, $\sigma$)} & \textbf{LCOM ($\mu$, $\sigma$)} & \textbf{WMC ($\mu$, $\sigma$)} \\
\midrule
Ant 1.7   & 11.04, 26.34 & 34.36, 36.02 & 89.14, 349.93 & 11.07, 11.97 \\
Camel 1.6 & 11.10, 22.52 & 21.20, 25.00 & 79.33, 523.75 & 8.57, 11.20 \\
\bottomrule
\end{tabular}%
}
\end{table}

\subsection{LLM Explanation Categories}
We explored the ability of LLMs to generate useful explanation for fault proneness metrics using ChatGPT-5.~\footnote{\url{https://chat.openai.com/}} 
We propose three kinds of explanations LLMs can provide to support developers' ability to understand and act on fault prediction metrics: \textbf{descriptive}, \textbf{contextual}, and \textbf{actionable}.

\textbf{Descriptive explanations}~\cite{acharya2023llm}  address foundational comprehension by providing plain-language definitions of what a metric fundamentally measures and why it matters for software quality. Johnson et al. \cite{johnson2013don,smith2015questions} documented that developers consistently ask definitional questions when encountering static analysis outputs ("What does this metric even mean?"), particularly when tool notifications use technical jargon or assume familiarity with software engineering concepts. This "comprehension barrier" is especially acute for OSS newcomers from diverse educational backgrounds \cite{steinmacher2015barriers,steinmacher2015social} who may lack formal training in object-oriented metrics. Cognitive load theory suggests that without foundational schemas, developers may experience high intrinsic cognitive load that impedes further reasoning \cite{sweller1988cognitive}. Descriptive explanations reduce this load by establishing basic mental models in an accessible language. By removing the prerequisite of specialized knowledge, descriptive explanations democratize access to fault prediction insights, enabling contributors across experience levels to build the foundational understanding necessary for informed decision-making.

\textbf{Contextual explanations}~\cite{lim2009and} answer the question \textit{what does this metric mean and why does it matter in the specific context of our project}.
This situates the metric within the project's own norms and history. It could include project-specific benchmarks (e.g. what is a typical CBO in this repository?), historical trends (has this metric been
rising for this component?), domain-specific considerations, or comparisons to similar projects.
Grattan et al. \cite{grattan2024need} found that 68\% of fault prediction models provide raw metric values without contextualization, forcing developers to independently determine whether a value represents a genuine anomaly or an architectural necessity. This gap is particularly problematic because identical metric values carry vastly different implications across projects due to varying architectural styles, coding conventions, and domain requirements\cite{johnson2016cross}.Contextual explanations address this by incorporating project-specific statistical baselines (mean, standard deviation,etc.) that ground qualitative assessments in quantitative evidence. For OSS newcomers who lack the project history to calibrate metric significance, contextual explanations reduce extraneous cognitive load by eliminating the need for extensive codebase exploration to establish norms.

\textbf{Actionable explanations}~\cite{siramgari2024raw} provides insights to \textit{what concrete steps could improve this metric and the underlying code} to turn understanding into action. 
This kind of explanation can suggest specific refactoring or redesign strategies, patterns to apply or avoid, and prioritize which changes would  have the most impact. 

 Smith et al. \cite{smith2015questions} identified this "knowing-doing gap" where developers recognize potential risks but struggle to formulate appropriate responses. Empirical studies reveal that developers frequently ask procedural questions when encountering static analysis warnings ("What should I actually do about this?"), yet tools typically provide detection without prescription \cite{johnson2013don}. This gap is especially challenging for OSS newcomers who lack the architectural experience that experts draw upon when planning refactoring or risk mitigation strategies. 
 Actionable explanations address this by providing specific tactics (e.g., design patterns, dependency management strategies), prioritization guidance (what to fix first), and preventive measures (how to avoid worsening metrics).

\begin{table*}[!t]
\centering
\caption{Comparing Explanation Taxonomies}
\label{tab:explanation_types}
\begin{tabularx}{\textwidth}{p{2.5cm} p{5cm} X}
\toprule
\textbf{Category} & \textbf{Purpose} & \textbf{Example} \\
\midrule
\textbf{Descriptive} & Foundational understanding & Metric definition, why it matters, calculation basis \\
\textbf{Contextual}  & Project-specific meaning   & Statistical benchmarks, outlier analysis, historical trends \\
\textbf{Actionable } & Possible next steps   & Refactoring patterns, priority guidance, testing strategies \\
\bottomrule
\end{tabularx}
\end{table*}

\subsection{LLM Prompt Composition}
To systematically evaluate the effectiveness of LLMs in generating contextually-aware fault metric explanations, we developed a structured prompt composition approach that incorporates project-specific statistical baselines. 
This approach addresses the limitation of generic explanations by providing LLMs with quantitative context about each project's characteristic metric distributions.

\subsubsection{Prompt Structure Design}
Our prompt composition follows a four component template designed to elicit comprehensive, project-contextualized explanations:

\textbf{Component 1: Project Context Establishment} We provide the LLM with the target project's name and baseline metric statistics including mean ($\mu$) and standard deviation ($\sigma$) for each Chidamber-Kemerer metric as derived from Table~\ref{tab:project-metric-stats}.

\textbf{Component 2: Class-Specific Metric Presentation} Individual class metrics are presented with their raw values, allowing the LLM to perform contextual analysis by comparing against the provided project baselines.

\textbf{Component 3: Structured Explanation Requirements} We explicitly request for explanation based on our defined categories:(1) definitional (2) contextual (3) actionable. 

\textbf{Component 4: Output Format Specification} We conclude the prompt with a formatting guidelines requesting developer-friendly language suitable for code review contexts.








\section{Illustrative Scenarios}
To determine the effectiveness of our context-aware LLM prompting methodology, we present two representative scenarios that showcase how our approach generates explanations across different metric severity levels (\textit{high}, \textit{low}). Each scenario includes the actual input prompt, the LLM-generated explanation, and our analysis of the output quality relative to our taxonomy requirements.



\subsection{Scenario 1: High-Risk Class}

\textbf{Class:} \texttt{Exchange.java} (Apache Camel 1.6)

\textbf{Class Profile:} Highest record of historical bugs 

\subsubsection{\textbf{LLM Input Prompt:} }

\begin{quote}
\small
``Explain the following software metrics for class \texttt{Exchange.java} to a new contributor to the Apache Camel project.\\
\textbf{\texttt{Exchange.java} class metrics:} CBO=448, RFC=26, LCOM=325, WMC=26\\
\textbf{Project Context:}Apache Camel 1.6 codebase with baseline statistics -- CBO: $\mu=11.10$, $\sigma=22.52$; RFC: $\mu=21.20$, $\sigma=25.00$; LCOM: $\mu=79.33$, $\sigma=523.75$; WMC: $\mu=8.57$, $\sigma=11.20$.

\textbf{Required Analysis:} (1) Clear definition of each metric and what it measures, (2) Analysis of what the \texttt{Exchange} class metrics indicate in this project context, (3) Actionable improvement suggestions based on the project baselines and \texttt{Exchange} class metrics. Use clear, actionable language suitable for code review discussions.'' 
\end{quote}

\subsubsection{LLM Generated Explanation (Key Excerpts)}
The LLM generated a comprehensive explanation addressing all required components. Key excerpts include:

\textbf{Metric Description:}
\begin{quote}
\small
\textit{``CBO — Coupling Between Object classes: How many distinct other classes Exchange touches(calls, references in fields/params/returns, inheritance, etc.), Higher = more dependencies and greater ripple risk when anything changes.''}
\end{quote}

\textbf{Contextual Analysis:}
\begin{quote}
\small
\textit{``CBO = 448 → Extreme outlier ($~19.4\sigma$ above the mean).
Interpretation: Exchange is a hub that depends on hundreds of types. Any change to Exchange or to many of those types risks widespread breakage.''}
\end{quote}

\textbf{Actionable Suggestions:}
\begin{quote}
\small
\textit{``Do not add new dependencies to Exchange unless absolutely unavoidable. Hide dependencies behind interfaces: if Exchange needs data from many types, depend on 1–2 facade/ports instead of dozens of concrete classes.''}
\end{quote}

\textit{[Complete LLM response available here: \cite{ArtifactsCodeRisk} ]}

\subsection{Scenario 2: Low-Risk Class}

\textbf{Class:} \texttt{DispatchTask.java} (Apache Ant 1.7)

\textbf{Class Profile:} No historical record of bugs.

\subsubsection{LLM Input Prompt:}
\begin{quote}
\small
``Explain the following software metrics for class \texttt{DispatchTask.java} to a new contributor to the Apache Ant 1.7 project.\\
\textbf{\texttt{DispatchTask.java} class metrics:} CBO=3, RFC=5, LCOM=4, WMC=4\\
\textbf{Project Context:}Apache Ant 1.7 codebase with baseline statistics -- CBO: $\mu=11.04$, $\sigma=26.34$; RFC: $\mu=34.36$, $\sigma=36.02$; LCOM: $\mu=89.14$, $\sigma=349.93$; WMC: $\mu=11.07$, $\sigma=11.97$.

\textbf{Required Analysis:} (1) Clear definition of each metric and what it measures, (2) Analysis of what the \texttt{DispatchTask} class metrics indicate in this project context, (3) Actionable improvement suggestions based on the project baselines and \texttt{DispatchTask} class metrics. Use clear, actionable language suitable for code review discussions.'' 
\end{quote}

\subsubsection{LLM Generated Explanation (Key Excerpts)}
The LLM generated a comprehensive explanation addressing all required components. Key excerpts include:

\textbf{Metric Description:}
\begin{quote}
``CBO (Coupling Between Objects): Number of other classes this class depends on; higher = more ripple risk.

RFC (Response For a Class): Methods in the class plus the distinct methods they call; higher = larger behavioral surface.

LCOM (Lack of Cohesion of Methods): Degree to which methods work on different state; higher = mixed responsibilities, lower cohesion.

WMC (Weighted Methods per Class): Sum of method complexities; higher = more or harder methods to read, test, and maintain.''
\end{quote}

\textbf{Contextual Analysis:}
\begin{quote}
\small
\textit{``DispatchTask is a small, cohesive task with low coupling (CBO) and a narrow behavioral surface (RFC). LCOM suggests methods operate on shared state, and WMC  indicates low per-method complexity. Within Ant 1.7, this profile aligns with a stable, non-hotspot class."}
\end{quote}

\textbf{Actionable Suggestions:}
\begin{quote}
\small
\textbf{Preserve low coupling:} avoid adding new concrete dependencies; prefer small interfaces.

\textbf{Keep the surface small: }resist new public methods; use collaborators for new behavior.

\textbf{Maintain cohesion:} if a change introduces an unrelated concern, extract a helper.

\textbf{Guard complexity:}split any growing conditional into intention-revealing helpers and add tests

\end{quote}

\textit{[Complete LLM response available here: \cite{ArtifactsCodeRisk} ]}

\section{Bridging the Explainability Gap}
Open source software development attracts contributors from diverse backgrounds, encompassing varying levels of technical expertise, cultural contexts, and language proficiencies~\cite{bosu2019diversity, guizani2022perceptions}. 
Recent research suggests LLMs can help reduce the cognitive burden that can comes with understanding and using technical documentation (e.g., as newcomers and non-native English speakers), indicating their potential to bridge accessibility gaps in technical communication and information sharing~ \cite{adejumo2024towards}.
Our proposed approach to leveraging LLMs for explainable automated fault prediction provides a foundation for research and practice in facilitating quality, metric-based open source contributions.


\subsection{Turning Metrics into Guidance}
Our illustrative scenarios showcasing the ability for LLMs to generate descriptive, contextual, and actionable explanations presents a promising signal for mitigating contribution risks in OSS environments. 
While program analysis tools, such as those measure fault proneness, can provide valuable information, research points to gaps in the ability for developers to fully comprehend and make use of the information provided without the necessary background or expertise~\cite{johnson2016cross,smith2015questions}.
By converting technical metrics into natural language guidance, we can reduce the cognitive load on contributors who may lack deep familiarity with either the analytical tools or the specific codebase (or both) thereby reducing the risk of low quality or risky contributions.

The democratization of technical insights has broader implications for how we design developer support systems, emphasizing the need for more collaborative tools that not only provide information, but guide solutions. 
Our early exploration suggests this approach could fundamentally alter the relationship between automated analysis and human developers. 
Instead of requiring contributors to interpret complex metrics independently, LLMs can serve as intelligent intermediaries that translate technical insights into actionable development guidance. This positions fault prediction not as a barrier to contribution, but as an enabler of more effective and confident code contributions across diverse developer populations.




\subsection{Aligning Risk with Project Baselines}

Along with indicating feasibility of our proposed approach, our efforts thus far underscore the importance of project-aware calibration in fault proneness assessment. 
Recent studies have indicated the value, and sometimes the necessity, of contextual insights when leveraging LLMs for software development tasks \cite{li2023large}.
A key insight from this work is that identical metric values can have vastly different implications across codebases due to varying project architectures, coding standards, and domain requirements. 
For example, when generating explanations \textcolor{nobaseline}{\textbf{without baseline metrics}} for \texttt{Exchange.java} we get vague assessments such ``extremely high'' and ``concerning'' (CBO=448) and metrics incorrectly flagged as problematic (e.g., ``moderate complexity'' for RFC=26).
While these may generally be considered concerning or problematic, making contributions based on these insights could lead to contributions that are outside the norms, expectation, or coding conventions of the project (all of which can impact software metrics across different types of projects)~\cite{mamun2017correlations}.
When generating with \textcolor{baseline}{\textbf{with baseline metrics}}, we get specific insights into metric values as it pertains to the project such as distance from mean for outliers (CBO=448 $\sim$19.4$\sigma$ above mean) and accurate interpretations (e.g., RFC=26 $\sim$+0.19$\sigma$ from baseline).
This emphasizes the importance of considering project-specific norms to contextualize metrics, and thereby explanations and guidance.







\subsection{Future Work}
Building on these foundations and insights, we plan to evaluate the effectiveness of LLM-generated metric explanations through a task-based user study, focusing on OSS contributors performing code maintenance tasks. The study will compare two conditions: (1) \textit{Baseline (Metrics-Only)} where participants are given the usual information that a tool might provide (e.g., raw metric scores) and (2) \textit{LLM-Augmented (Explanations)} where  participants have access to explanations generate by our proposed approach.
In both scenarios, users will be expected to make decisions on a course of action based on the information provided (e.g., which file to refactor or how to fix a code smell).
We will use realistic development scenarios (e.g., improving flagged modules, reviewing pull requests with automated analysis) to curate practical insights. We will measure three key outcomes: (1) Decision Quality - whether participants correctly identify risky code and choose appropriate actions, assessed by expert judges; (2) Time to Completion - testing our hypothesis that LLM explanations accelerate comprehension and decision-making by reducing time spent interpreting metrics; and (3) Error Rates - tracking whether participants make fewer wrong assumptions or misdirected efforts, as we expect explanations to reduce metric misinterpretation and analysis paralysis compared to baseline conditions. We will also solicit feedback that we can leverage to improve our approach and its potential for impact.
These efforts will provide a novel and valuable foundation for supporting risk assessment and mitigation in open source software development.


\bibliography{references}

\begin{thebibliography}{10}
\providecommand{\url}[1]{#1}
\csname url@samestyle\endcsname
\providecommand{\newblock}{\relax}
\providecommand{\bibinfo}[2]{#2}
\providecommand{\BIBentrySTDinterwordspacing}{\spaceskip=0pt\relax}
\providecommand{\BIBentryALTinterwordstretchfactor}{4}
\providecommand{\BIBentryALTinterwordspacing}{\spaceskip=\fontdimen2\font plus
\BIBentryALTinterwordstretchfactor\fontdimen3\font minus \fontdimen4\font\relax}
\providecommand{\BIBforeignlanguage}[2]{{%
\expandafter\ifx\csname l@#1\endcsname\relax
\typeout{** WARNING: IEEEtran.bst: No hyphenation pattern has been}%
\typeout{** loaded for the language `#1'. Using the pattern for}%
\typeout{** the default language instead.}%
\else
\language=\csname l@#1\endcsname
\fi
#2}}
\providecommand{\BIBdecl}{\relax}
\BIBdecl

\bibitem{steinmacher2015systematic}
I.~Steinmacher, M.~A.~G. Silva, M.~A. Gerosa, and D.~F. Redmiles, ``A systematic literature review on the barriers faced by newcomers to open source software projects,'' \emph{Information and Software Technology}, vol.~59, pp. 67--85, 2015.

\bibitem{steinmacher2015social}
I.~Steinmacher, T.~Conte, M.~A. Gerosa, and D.~Redmiles, ``Social barriers faced by newcomers placing their first contribution in open source software projects,'' in \emph{Proceedings of the 18th ACM conference on Computer supported cooperative work \& social computing}, 2015, pp. 1379--1392.

\bibitem{chidamber1994metrics}
S.~R. Chidamber and C.~F. Kemerer, ``A metrics suite for object oriented design,'' \emph{IEEE Transactions on software engineering}, vol.~20, no.~6, pp. 476--493, 1994.

\bibitem{denaro2002empirical}
G.~Denaro and M.~Pezze, ``An empirical evaluation of fault-proneness models,'' in \emph{Proceedings of the 24th International Conference on Software Engineering}, 2002, pp. 241--251.

\bibitem{subramanyam2003empirical}
R.~Subramanyam and M.~S. Krishnan, ``Empirical analysis of ck metrics for object-oriented design complexity: Implications for software defects,'' \emph{IEEE Transactions on software engineering}, vol.~29, no.~4, pp. 297--310, 2003.

\bibitem{balasubramaniam2022software}
S.~Balasubramaniam and S.~G. Gollagi, ``Software defect prediction via optimal trained convolutional neural network,'' \emph{Advances in Engineering Software}, vol. 169, p. 103138, 2022.

\bibitem{goyal2022effective}
S.~Goyal, ``Effective software defect prediction using support vector machines (svms),'' \emph{International Journal of System Assurance Engineering and Management}, vol.~13, no.~2, pp. 681--696, 2022.

\bibitem{grattan2024need}
N.~Grattan, D.~A. da~Costa, and N.~Stanger, ``The need for more informative defect prediction: A systematic literature review,'' \emph{Information and software technology}, vol. 171, p. 107456, 2024.

\bibitem{johnson2013don}
B.~Johnson, Y.~Song, E.~Murphy-Hill, and R.~Bowdidge, ``Why don't software developers use static analysis tools to find bugs?'' in \emph{2013 35th International Conference on Software Engineering (ICSE)}.\hskip 1em plus 0.5em minus 0.4em\relax IEEE, 2013, pp. 672--681.

\bibitem{dam2018explainable}
H.~K. Dam, T.~Tran, and A.~Ghose, ``Explainable software analytics,'' in \emph{Proceedings of the 40th international conference on software engineering: New ideas and emerging results}, 2018, pp. 53--56.

\bibitem{ahmad2021transformer}
W.~U. Ahmad \emph{et~al.}, ``Transformer models for code summarization,'' \emph{arXiv preprint arXiv:2107.05273}, 2021.

\bibitem{adejumo2024towards}
E.~K. Adejumo and B.~Johnson, ``Towards leveraging llms for reducing open source onboarding information overload,'' in \emph{Proceedings of the 39th IEEE/ACM International Conference on Automated Software Engineering}, 2024, pp. 2210--2214.

\bibitem{adejumo2025bridging}
E.~K. Adejumo, B.~Johnson, and M.~Guizani, ``Bridging language gaps in open-source documentation with large-language-model translation,'' \emph{arXiv preprint arXiv:2508.02497}, 2025.

\bibitem{tufano2024unveiling}
R.~Tufano, A.~Mastropaolo, F.~Pepe, O.~Dabic, M.~Di~Penta, and G.~Bavota, ``Unveiling chatgpt's usage in open source projects: A mining-based study,'' in \emph{Proceedings of the 21st International Conference on Mining Software Repositories}, 2024, pp. 571--583.

\bibitem{kang2024quantitative}
S.~Kang, G.~An, and S.~Yoo, ``A quantitative and qualitative evaluation of llm-based explainable fault localization,'' \emph{Proceedings of the ACM on Software Engineering}, vol.~1, no. FSE, pp. 1424--1446, 2024.

\bibitem{nam2024using}
D.~Nam, A.~Macvean, V.~Hellendoorn, B.~Vasilescu, and B.~Myers, ``Using an llm to help with code understanding,'' in \emph{Proceedings of the IEEE/ACM 46th International Conference on Software Engineering}, 2024, pp. 1--13.

\bibitem{ArtifactsCodeRisk}
\BIBentryALTinterwordspacing
E.~Adejumo, ``Artifacts, prompts and outputs,'' 2025. [Online]. Available: \url{https://github.com/INSPIRED-GMU/ExplainingCodeRisk}
\BIBentrySTDinterwordspacing

\bibitem{aggarwal2021software}
\BIBentryALTinterwordspacing
D.~Aggarwal, ``Software defect prediction dataset,'' Dataset, 2021. [Online]. Available: \url{https://doi.org/10.6084/m9.figshare.13536506.v1}
\BIBentrySTDinterwordspacing

\bibitem{basili2002validation}
V.~R. Basili, L.~C. Briand, and W.~L. Melo, ``A validation of object-oriented design metrics as quality indicators,'' \emph{IEEE Transactions on software engineering}, vol.~22, no.~10, pp. 751--761, 2002.

\bibitem{acharya2023llm}
A.~Acharya, B.~Singh, and N.~Onoe, ``Llm based generation of item-description for recommendation system,'' in \emph{Proceedings of the 17th ACM conference on recommender systems}, 2023, pp. 1204--1207.

\bibitem{smith2015questions}
J.~Smith, B.~Johnson, E.~Murphy-Hill, B.~Chu, and H.~R. Lipford, ``Questions developers ask while diagnosing potential security vulnerabilities with static analysis,'' in \emph{Proceedings of the 2015 10th Joint Meeting on Foundations of Software Engineering}, 2015, pp. 248--259.

\bibitem{steinmacher2015barriers}
I.~Steinmacher, M.~A.~G. Silva, M.~A. Gerosa, and D.~F. Redmiles, ``Barriers faced by newcomers to open source software projects,'' in \emph{Proceedings of the 18th ACM conference on Computer supported cooperative work \& social computing}, 2015.

\bibitem{sweller1988cognitive}
J.~Sweller, ``Cognitive load during problem solving: Effects on learning,'' \emph{Cognitive science}, vol.~12, no.~2, pp. 257--285, 1988.

\bibitem{lim2009and}
B.~Y. Lim, A.~K. Dey, and D.~Avrahami, ``Why and why not explanations improve the intelligibility of context-aware intelligent systems,'' in \emph{Proceedings of the SIGCHI conference on human factors in computing systems}, 2009, pp. 2119--2128.

\bibitem{johnson2016cross}
B.~Johnson, R.~Pandita, J.~Smith, D.~Ford, S.~Elder, E.~Murphy-Hill, S.~Heckman, and C.~Sadowski, ``A cross-tool communication study on program analysis tool notifications,'' in \emph{Proceedings of the 2016 24th ACM SIGSOFT International Symposium on Foundations of Software Engineering}, 2016, pp. 73--84.

\bibitem{siramgari2024raw}
D.~Siramgari and V.~Sikha, ``From raw data to actionable insights: Leveraging llms for automation. zenodo,'' 2024.

\bibitem{bosu2019diversity}
A.~Bosu and K.~Z. Sultana, ``Diversity and inclusion in open source software (oss) projects: Where do we stand?'' in \emph{2019 ACM/IEEE International Symposium on Empirical Software Engineering and Measurement (ESEM)}.\hskip 1em plus 0.5em minus 0.4em\relax IEEE, 2019, pp. 1--11.

\bibitem{guizani2022perceptions}
M.~Guizani, B.~Trinkenreich, A.~A. Castro-Guzman, I.~Steinmacher, M.~Gerosa, and A.~Sarma, ``Perceptions of the state of d\&i and d\&i initiative in the asf,'' in \emph{Proceedings of the 2022 ACM/IEEE 44th International Conference on Software Engineering: Software Engineering in Society}, 2022, pp. 130--142.

\bibitem{li2023large}
J.~Li, C.~Tao, J.~Li, G.~Li, Z.~Jin, H.~Zhang, Z.~Fang, and F.~Liu, ``Large language model-aware in-context learning for code generation,'' \emph{ACM Transactions on Software Engineering and Methodology}, 2023.

\bibitem{mamun2017correlations}
M.~A.~A. Mamun, C.~Berger, and J.~Hansson, ``Correlations of software code metrics: an empirical study,'' in \emph{Proceedings of the 27th international workshop on software measurement and 12th international conference on software process and product measurement}, 2017, pp. 255--266.

\end{thebibliography}

\end{document}